\documentclass[preprint,preprintnumbers,nofootinbib,aps,14pt,onecolumn]{revtex4-1}
\usepackage{amsmath,amssymb,bm}
\usepackage{graphicx}
\usepackage{hyperref} 
\newcommand{\gs}{\gamma_{S}}

\newcommand{\pipm}{\pi^{\pm}}
\newcommand{\ppbar}{\ensuremath{\mathrm {p(\bar{p})}}}
\newcommand{\xixibar}{\Xi^{-}(\bar{\Xi}^{+})}
\newcommand{\kapm}{\rm{K}^{\pm}}
\newcommand{\omombar}{\Omega(\bar{\Omega})}

\newcommand{\kpi}{(K^{+}+K^{-})/(\pi^{+}+\pi^{-})}
\newcommand{\kspi}{2\rm{K}_{S}^{0}/(\pi^{+}+\pi^{-})}
\newcommand{\ppi}{(\mathrm{p}+\bar{\mathrm{p}})/(\pi^{+}+\pi^{-})}
\newcommand{\llpi}{(\mathrm{\Lambda}+\bar{\mathrm{\Lambda}})/(\pi^{+}+\pi^{-})}
\newcommand{\xipi}{(\mathrm{\Xi}+\bar{\mathrm{\Xi}})/(\pi^{+}+\pi^{-})}
\newcommand{\omegapi}{(\mathrm{\Omega}+\bar{\mathrm{\Omega}})/(\pi^{+}+\pi^{-})}
\newcommand{\llks}{(\mathrm{\Lambda}+\bar{\mathrm{\Lambda}})/2\rm{K}_{S}^{0}}

\newcommand{\pp}{\ensuremath{\mathrm {p\kern-0.05em p}}} 
\newcommand{\PbPb}{\ensuremath{\mbox{Pb--Pb}}} 
\newcommand{\pPb}{\ensuremath{\mbox{p--Pb}}} 

\newcommand{\mtpi}{\langle m_{T}^{\pi} \rangle}
\newcommand{\pt}{\langle p_{T}^i \rangle}
\newcommand{\dndy}{\textrm{d}N_{ch}/\textrm{d}y}
\newcommand{\dndymid}{\textrm{d}N_{ch}/\textrm{d}y|_{|y|\leq 0.5}}

\newcommand{\dndetamid}{\textrm{d}N_{ch}/\textrm{d}\eta|_{|\eta|\leq 0.5}}
\newcommand{\dndetamidone}{\frac{\textrm{d}N_{ch1}}{\textrm{d}\eta}|_{|\eta|\leq 0.5}}
\newcommand{\ddndetamidone}{\frac{\textrm{d}^{2}N_{ch1}}{\textrm{d}\eta}|_{|\eta|\leq 0.5}}

\begin{document}
\title {The Effect of Multiplicity Cut on Particle Ratios at 7 TeV Proton-Proton Collisions}
\author{Ehab~Abbas\footnote{Email: ehab.g.abbas@gmail.com}}
\affiliation{Physics Department, Faculty of Education, Ain Shams University, Cairo, Egypt.}
\date{\today}

\begin{abstract}
The final state hadronic composition of proton-proton collisions at 7 TeV as a function of charged particle multiplicity density is studied. The thermal model is adjusted to match the experimental conditions in order to understand the effect of the experimental cuts on rapidity and multiplicity. This study indicates that the measured hadronic composition under a simultaneous cut on rapidity and multiplicity does not coincide with the fireball hadronic composition. Excluding proton to pion ratio, the adjusted model can explain the measured particle ratios. Therefore, the observed strangeness enhancement as a function of charged particle multiplicity density could be an effect of the experimental cuts.
\end{abstract}. 


\maketitle


\section{Introduction}
\label{sec:intro}
      An enhancement in (multi-)strange particles production, relative to pions, measured at pseudo-rapidity $|\eta| \leq 0.5$ in proton-proton (pp) collisions at center of mass energy, $\sqrt{s}$,= 7 TeV, is reported by ALICE collaboration as a function of charged particle multiplicity density, $\dndetamid$ \cite{ALICE:2017jyt}.
The enhancement of multi-strange baryons is more pronounced than single strange hadrons.
This data \cite{ALICE:2017jyt} is very interesting because it links results of different colliding systems and energies by suggesting that the hadronic composition strictly depends on the number of final state particles.
Furthermore, the hadron production in high-multiplicity pp events seems to resemble the one in heavy ion collisions.
       
The statistical hadronization model (SHM) \cite{BraunMunzinger:2003zd} has successfully described the final state hadronic composition for pp at low \cite{Becattini:1996gy} and LHC energies \cite{Das:2016muc}, without any multiplicity cut. 
However, the model success is almost fading when it used to describe the measured data at different $\dndetamid$ \cite{Vislavicius:2016rwi,Sharma:2018uma}. 
The application of canonical ensemble with exact baryon, strangeness and charge conservation to this data, is questionable because charges do not have to be exactly conserved at limited phase space. 
Furthermore, the canonical ensemble with exact strangeness conservation, which supposed to suppress the strangeness particles at low $\dndetamid$, needs strangeness suppression factor, $\gs$, greater than one to fit the data \cite{Sharma:2018uma}. 
In Ref.\cite{Vislavicius:2016rwi}, the same ensemble was used under the assumption of full chemical equilibrium ($\gs = 1$) and it leads to qualitatively description of the enhancement \cite{ALICE:2017jyt} but it fails to explain multiplicity dependence of $\Phi/\pi$ in proton-Lead ($\pPb$). 
Although grand canonical is the most appropriate ensemble in the considered case, it does not show an acceptable agreement with the data \cite{ALICE:2017jyt} where $\chi^2/Ndf$ (fit quality) varies from 3 to 10 as $\dndetamid$ increases \cite{Sharma:2018uma}. 
      
Since SHM can not satisfactory reproduce the experimental ratios \cite{ALICE:2017jyt}, it is convenient to turn our attention to missing physical processes or mismatch between the experimental conditions and the model.
In the present work, the model was adjusted to match the experimental conditions, the cuts on $\dndetamid$ and rapidity, which enable us to answer whether the measured particle ratios under these cuts represent the fireball hadronic composition or not. The cut on multiplicity is interpreted, in this study, to a cut on fireballs-rapidities which has an important influence on the particle ratios at midrapidity. The paper is organized as follows. Sec.(\ref{sec:model}) elaborates details about the model adjustment. Sec.(\ref{sec:results}) is devoted to present the results and discussion. The conclusions and outlook are presented in Sec.(\ref{sec:con}).
\section{MODEL}
\label{sec:model} 
 When the two colliding proton overlap, a fraction of incident energy is converted into particles where the residual energy is still carried by fireball as longitudinal momentum.
Since longitudinal momentum of the leading hadrons are uncorrelated \cite{Basile:1982we}, it is convenient to deal the hadronization of the two fireballs separately.
Due to the thermal features of particle production in pp collisions \cite{Becattini:1996gy}, hadronization will be treated thermally.
 The two fireballs, right-moving and left-moving, are assumed to be distributed uniformly along the rapidity axis, as a crude guess. 
The fireball rapidity, $Y_{FB}$, ranges from 0 to $Y_{FB}^{inel}$, the maximum fireball rapidity, which is calculated as
\begin{equation} 
\label{Y_FB_inel}
           Y_{FB}^{inel}  = cosh^{-1} \Big(\frac{0.5\sqrt{s_{NN}}}{(m_p+\left\langle m_T^{\pi} \right\rangle)} \Big),
\end{equation}
where $m_p$  is proton mass and $\mtpi$ is pion average transverse mass. 
The contribution of any fireball in particle production, is the multiplication of number of charged particles emitted by the fireball, $N_{ch}(Y_{FB})$, by its probability, $\psi(Y_{FB})$.

   According to SHM, about 61.6\% of the final state particles -after resonances decay- are charged \cite{Tawfik:2013bza}.
 The experimental yields, measured at mid-rapidity \cite{Adam:2015qaa,Abelev:2012jp} were used to calculate the probabilities, $P$, of having a charged particle as $\pipm$, $\kapm$, $\ppbar$, $\xixibar$ or $\omombar$ which are 0.84301, 0.10739, 0.046562, 0.002928 and 0.0002534, respectively.
These ratios are taken to be the fireball hadronic composition as long as they are rapidity independent over a relative large interval around midrapidity \cite{Becattini:2003wp}. The measured antibaryon/baryon ratios, from mid-rapidity up to three units of rapidity, exhibit almost no change \cite{Abbas:2013rua} which means that the cut at midrapidity has no effect and the measured ratios still represent fireball hadronic composition \cite{Becattini:2003wp}.
 
\begin{figure}[htb]
\centering{
\includegraphics[scale=0.6]{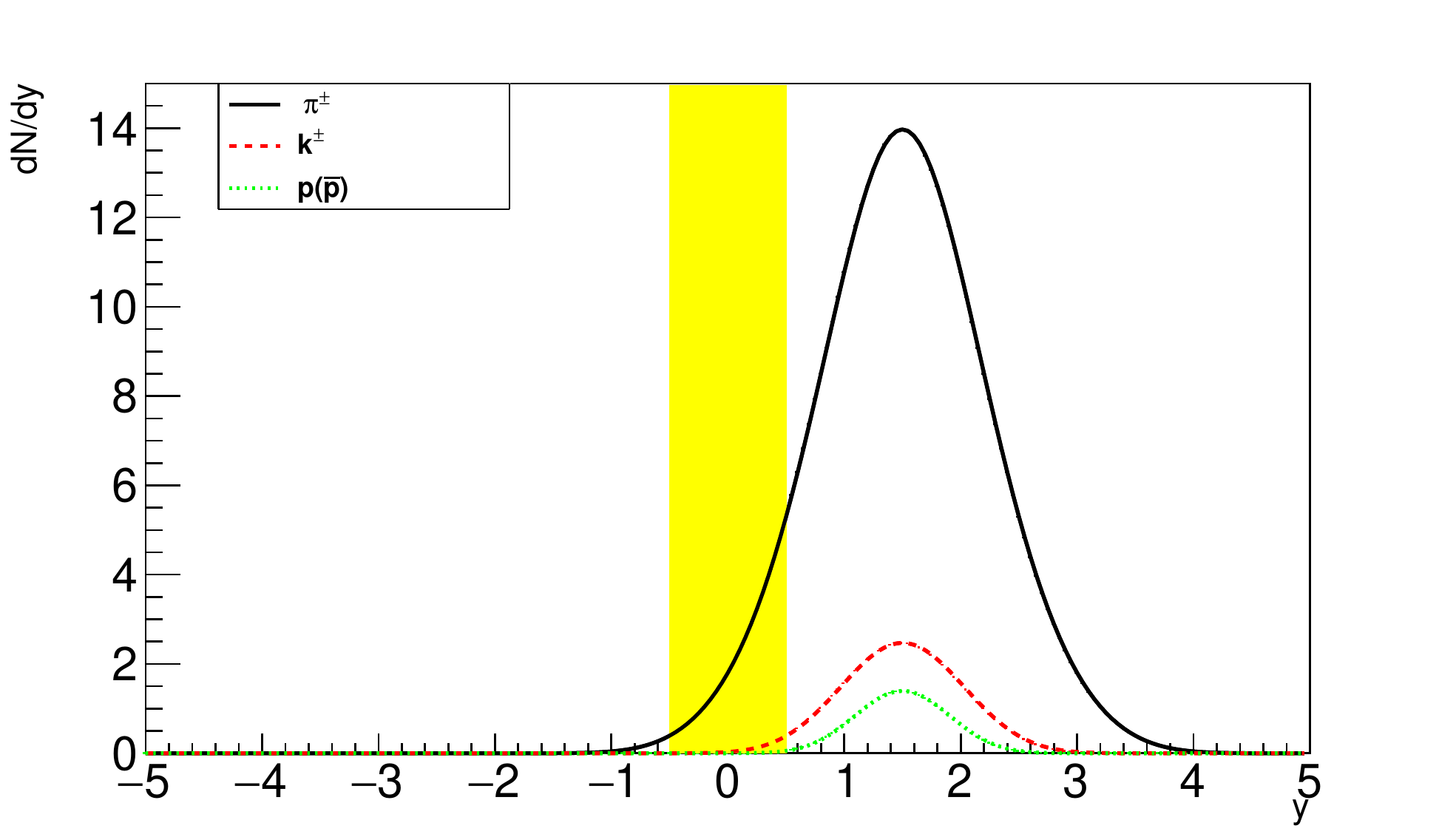}
\caption{The rapidity distributions for different particle species emitted form fireball moving with $Y_{FB} = 1.5$ where the yellow band represents the measured region for the considered data.} \label{fig:fig1} } 
\end{figure}

    The rapidity distribution of charged particles, $\dndy$, ignoring $\xixibar$ and $\omombar$ contributions, from one fireball can be obtained as 
\begin{eqnarray}
\frac{dN_{ch}}{dy} &=& \frac{dN_{\pi^{\pm}}}{dy} + \frac{dN_{k^{\pm}}}{dy}+ \frac{dN_{\ppbar}}{dy} \\
&=&N_{ch}(Y_{FB}) (P_{\pi^{\pm}}\frac{dn_{\pi}}{dy} + P_{k^{\pm}}\frac{dn_{k}}{dy}+ P_{\ppbar} \frac{dn_{p}}{dy}),
\label{eq:dny}
\end{eqnarray} 
where $\frac{dn_{i}}{dy}$ is the rapidity distribution of one particle species $i$ radiated from fireball moving with rapidity $Y_{FB}$, which is given as  
\begin{equation}
\label{eq:d1dy}
\frac{dn_{i}}{dy} = \frac{1}{2 K_2(m_{i}/T)}  
        \Big({1  + \frac{2T}{m_{i} cosh(y-Y_{FB})}+  \frac{2T^2}{m_{i}^2 cosh^2(y-Y_{FB})}} \Big)
        \exp\left(\frac{-m_{i} cosh(y-Y_{FB})}{T} \right),
\end{equation}
where $K_2(m_{i}/T)$ is the 2nd order modified Bessel function,
      $m_{i}$ is the particle species i mass and 
      $T$ is the temperature taken to be 0.163 GeV\cite{ALICE:2017jyt}. 
The rapidity distributions of $\pipm$, $\kapm$ and $\ppbar$ originated from a fireball moving with $Y_{FB}= 1.5$ and emitting 30 charged particles are shown in Fig.(\ref{fig:fig1}). 
The effect of the resonances was implicitly included in $P_{i}$ values, however they have negligible effect on rapidity distributions shape \cite{Schnedermann:1993ws}. 
Fig.(\ref{fig:fig1}) illustrates that particle ratios at the mid-rapidity, changes with fireball rapidity. 
When the fireball moves with high rapidity, a suppression of heavy particle mass yields compared to the light ones is expected at the measured region and vice verse.
On the other hand, Fig.(\ref{fig:fig1}) also demonstrates the dependence of $\dndymid$ on the fireball rapidity. 
As the fireball rapidity increases, the $\dndymid$ decreases.
In view of that, an artificial enhancement, in heavy particles with respect to pions produced at midrapidity, is excepted as $\dndymid$ increases.     

\begin{figure}[htb]
\centering{
\includegraphics[scale=0.43]{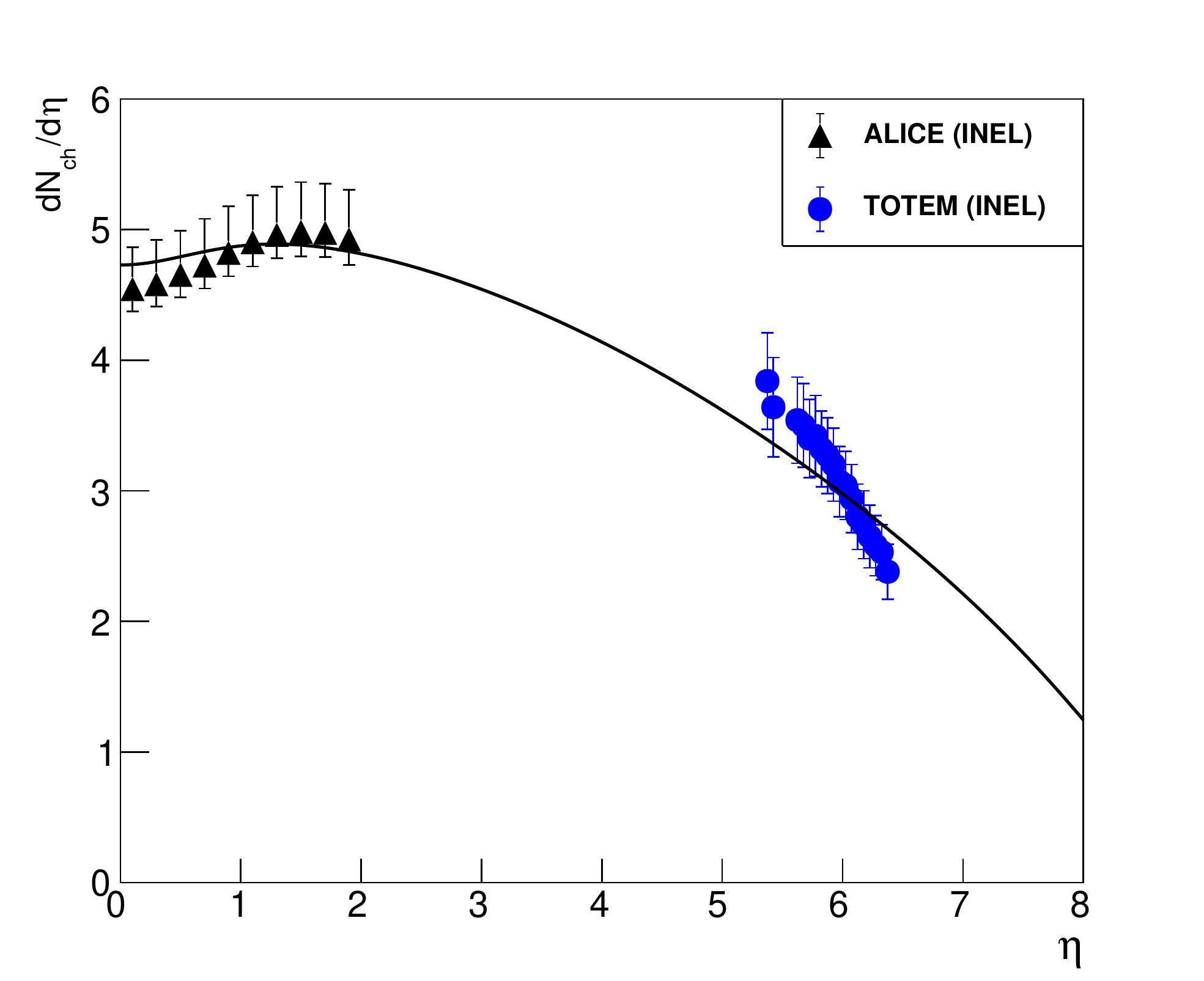}
\includegraphics[scale=0.43]{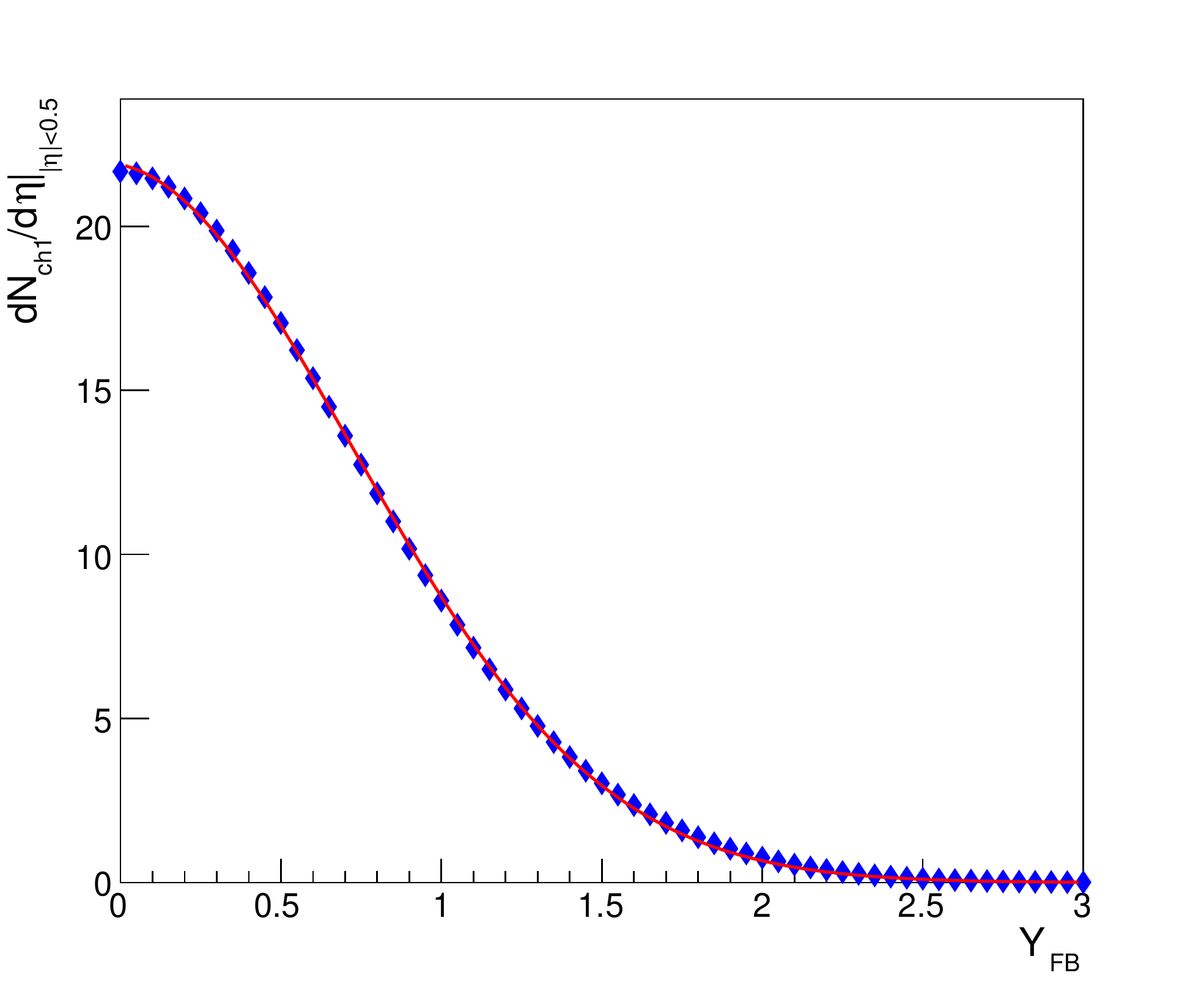}
\caption{Left: The experimental pseudorapidity distribution of charged particles \cite{Aspell:2012ux,Adam:2015gka} (symbols) is compared to the calculations (line) from two fireballs at 7 TeV pp collisions, which assure minimum $\chi^2$. Right: The fireball-rapidity dependence of $\dndetamidone$ calculated by Eq.(\ref{eq:dny},\ref{eq:eta})(symbles) and fitted with Eq.(\ref{eq:main})(line).} \label{fig:fig2} } 
\end{figure}

The pseudo-rapidity distribution of charged particles can be obtained by multiplying Eq.(\ref{eq:dny}) by the Jacobian,
\begin{equation}
 \label{eq:eta}
 J(y,m_i) = \sqrt{1-\frac{m_i^2}{ (m_i^2+\pt^2)cosh^2({y})}},
\end{equation}
where $\pt$ denotes average transverse momentum of particle species $i$, taken from Ref.\cite{Adam:2015qaa}. A relation between the number of charged particle produced at $|\eta| \leq 0.5$ from one fireball, $\dndetamidone$, and $Y_{FB}$ can be established using Eq.(\ref{eq:dny},\ref{eq:eta}) where $N_{ch}(Y_{FB})$ can be extracted by fitting the measured pseudo-rapidity distribution  of charged particle \cite{Aspell:2012ux,Adam:2015gka} with the corresponding theoretical one.
The pseudo-rapidity distribution of charged particles from right-moving fireballs reads
\begin{equation} 
\label{eq:detafit}
\frac{dN_{ch}}{d\eta}= 
\int_0^{Y_{FB}^{inel}} ~\psi(Y_{FB})
 N_{ch}(Y_{FB}) ~\frac{dn_{ch}}{d\eta}( Y_{FB}) ~ d Y_{FB}.
\end{equation}
where $dn_{ch}/d\eta$ is the pseudo-rapidity distribution from one charged particle emitted from fireball moving with rapidity $Y_{FB}$. 
The integration represents the contribution of all fireballs at a certain pseudo-rapidity, $\eta$. $N_{ch}(Y_{FB})$ is assumed to be in this form $N_{ch}^0(1+nY_{FB}^2)$ where $N_{ch}^0$ and $n$ are free parameters.
This form adapted a slight deviation from the assumption of boost-invariant fireballs contribution \cite{Schnedermann:1993ws}.
The fit is performed by reaching $\chi^2$ minimum, as shown Fig.(\ref{fig:fig2})(left), yielding $N_{ch}^0$ =  $43.4605$ and $n$ = $-0.011393$ 
with $\chi^2/Ndf = 0.5346$.

Using $N_{ch}(Y_{FB})$, $\dndetamidone$ can be calculated using Eq.(\ref{eq:dny},\ref{eq:eta}), as shown in Fig.(\ref{fig:fig2})(right). The calculated results can be expressed by a Gaussian form,
\begin{equation} 
\label{eq:main}
 \dndetamidone = C_{0} e^{-0.5(\frac{Y_{FB}-C_{1}}{C_{2}})^2}
\end{equation} 
where  $C_{0}$, $C_{1}$ and $C_{2}$ are $21.986$, $-0.0606016$ and $0.779198$, respectively. Consequently,  $Y_{FB}$ can be converted to $\dndetamidone$ and vice versa. The dependence shown in Fig.(\ref{fig:fig2})(right), is due to two reasons. The first is that the number of produced charged particles, reflecting the deposited energy, decreases as the fireball rapidity increases. Secondly, the fireballs moving with low-rapidity, suffered a deacceleration which makes the radiated particles’ rapidities close to midrapidity.

Henceforth, we deal with two fireballs to find the rapidity pairs ($Y_{FB2}$, $Y_{FB1}$) which lead to create a certain number of charged particle at mid-pseudorapidity from both, $\dndetamid$.
Each pair has a probability and the sum of all these probability should give the total probability for multiplicity class, as shown by this equation 
 \begin{equation} 
 \sigma_{X} = \int_{0}^{X}
   \textstyle \psi_{1}\scriptstyle( \dndetamidone)~~
   \textstyle \psi_{2}\scriptstyle( X - \dndetamidone) 
            ~~\ddndetamidone
\end{equation}
where $\psi_{1}$ and $\psi_{2}$ are the probabilities for the two fireball and X is the value of charged particle multiplicity density, $\dndetamid$.
The calculations presented in this work, were done numerically via replacing the integration by a sum
\footnote{X was divided into bins with size 0.01}
as follow
\begin{equation} 
 \sigma_{X} = \sum_{\scriptstyle \dndetamidone = 0}^{X} 
 \textstyle \psi_{1}\scriptstyle(\dndetamidone)~~ 
 \textstyle \psi_{2}\scriptstyle(X - \dndetamidone).           
\end{equation} 
 Consequently, the particle yields $j$ calculated under the two cuts reads
\begin{multline}
\label{eq:yields}
j_{X}  = \hspace{7mm} \frac{1}{\sigma_{X}} \sum_{\scriptstyle \dndetamidone = 0}^{X} \textstyle \psi_{1}\scriptstyle( \dndetamidone)~~ \textstyle \psi_{2}\scriptstyle( X - \dndetamidone) \\ \textstyle
               [N_{ch}(Y_{FB1}) P_{j} \int_{-0.5}^{0.5} \frac{dn_{j}}{dy}(Y_{FB1})~dy+
                N_{ch}(Y_{FB2}) P_{j} \int_{-0.5}^{0.5} \frac{dn_{j}}{dy}(Y_{FB2})~dy],
\end{multline}  
where $Y_{FB1}$ and  $Y_{FB2}$ correspond to $\dndetamidone$ and $ X -\dndetamidone$, respectively, as described in Eq.(\ref{eq:main}). In Eq.(\ref{eq:yields}), the two terms between brackets represent the number of $j$ particles produced by both fireballs moving with rapidities that make charged particle multiplicity density from both equals to X. Thus,  the ratio between any two particle yields $j$ and $i$ can be written as 
\begin{multline} 
\label{eq:ratio1}
 \hspace{7mm}  (\frac{j}{i})_{X} = (\frac{j}{i})_{(fireball)} \\
 \frac
 {\sum_{\scriptstyle \dndetamidone = 0}^{X} \psi_{1}\scriptstyle(\dndetamidone)\textstyle \psi_{2}\scriptstyle(X - \dndetamidone) \textstyle
   [N_{ch}(Y_{FB1}) \int_{-0.5}^{0.5} \frac{dn_{j}}{dy} (Y_{FB1})dy+
    N_{ch}(Y_{FB2}) \int_{-0.5}^{0.5} \frac{dn_{j}}{dy} (Y_{FB2})dy]}
 {\sum_{\scriptstyle \dndetamidone = 0}^{X} \psi_{1} \scriptstyle(\dndetamidone)\textstyle \psi_{2}       
  \scriptstyle (X - \dndetamidone) \textstyle
   [N_{ch}(Y_{FB1}) \int_{-0.5}^{0.5} \frac{dn_{i}}{dy} (Y_{FB1})dy+
    N_{ch}(Y_{FB2}) \int_{-0.5}^{0.5} \frac{dn_{i}}{dy} (Y_{FB2})dy]}.
\end{multline}
This equation forms a relation between the particle ratios with/without cuts where the big term stands for the cut modification factor as function of $Y_{F B}$. When this term doesnot equal one this means that measured hadronic composition under both cuts does not coincide with the fireball hadronic composition.
\section{Results and Discussion}
\label{sec:results}
\begin{figure}[htb]
\centering{
\includegraphics[scale=0.83]{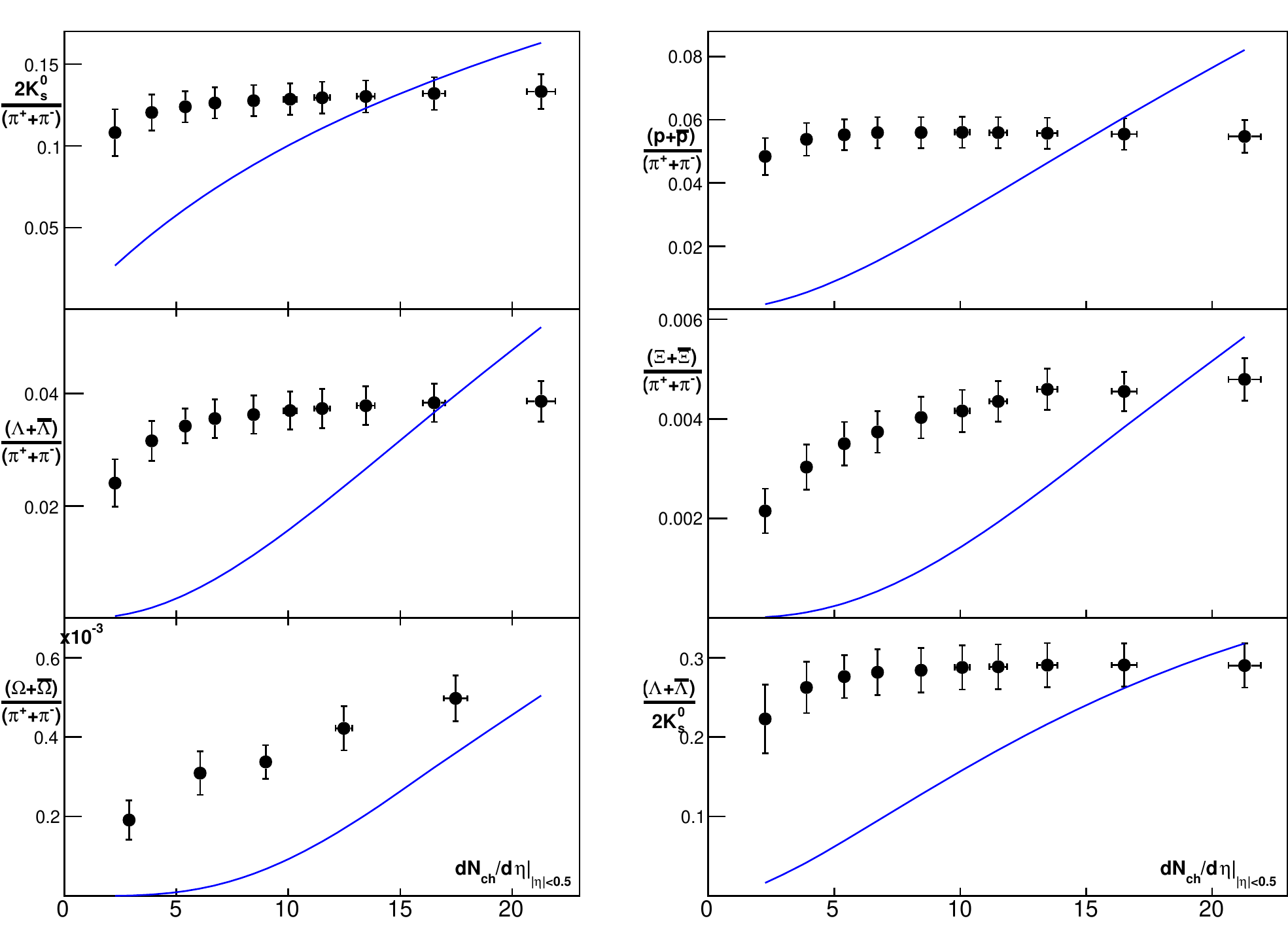}
\caption{ Different particle ratios as a function of $\dndetamid$ calculated from  Eq.(\ref{eq:ratio1}) (line) versus the experimental ones (symbols) \cite{ALICE:2017jyt}} \label{fig:fig3} } 
\end{figure}

The theoretical ratios are calculated according to Eq.(\ref{eq:ratio1}) taking into consideration the charged particle multiplicity densities from both fireballs to match the experimental multiplicity cut for proper comparison. 
The fireball ratios, in Eq.(\ref{eq:ratio1}), are 0.1274, 0.055011, 0.003519 and 0.000301 for $\kspi$\footnote{The experimental value of $\kpi$ was taken to be $\kspi$.}, $\ppi$, $\xipi$ and $\omegapi$, respectively measured under no multiplicity cut \cite{Adam:2015qaa,Abelev:2012jp}, where $\llpi$ was calculated to be 0.0334 using SHM \cite{Tawfik:2013bza} at zero-net baryon density and 0.1585 GeV chemical freeze-out temperature.
These ratios were implemented to calculate the ones under the multiplicity cut, as shown in Fig.(\ref{fig:fig3}).
The lines in this figure represent the calculations under the cuts effect, Eq.(\ref{eq:ratio1}), assuming that all fireballs have the same chemical composition.  The effect of both cuts lead to an artificial enhancement in heavy mass particles with respect to light mass ones.
The effect of the cuts becomes more pronounced for massive particles. The lack information about  $\psi(Y_{FB})$ and $N(Y_{FB})$, makes the comparison with the experimental data in Fig.(\ref{fig:fig3}) has no value except demonstrating the experimental cuts effect.

\begin{figure}[htb]
\centering{
\includegraphics[scale=0.83]{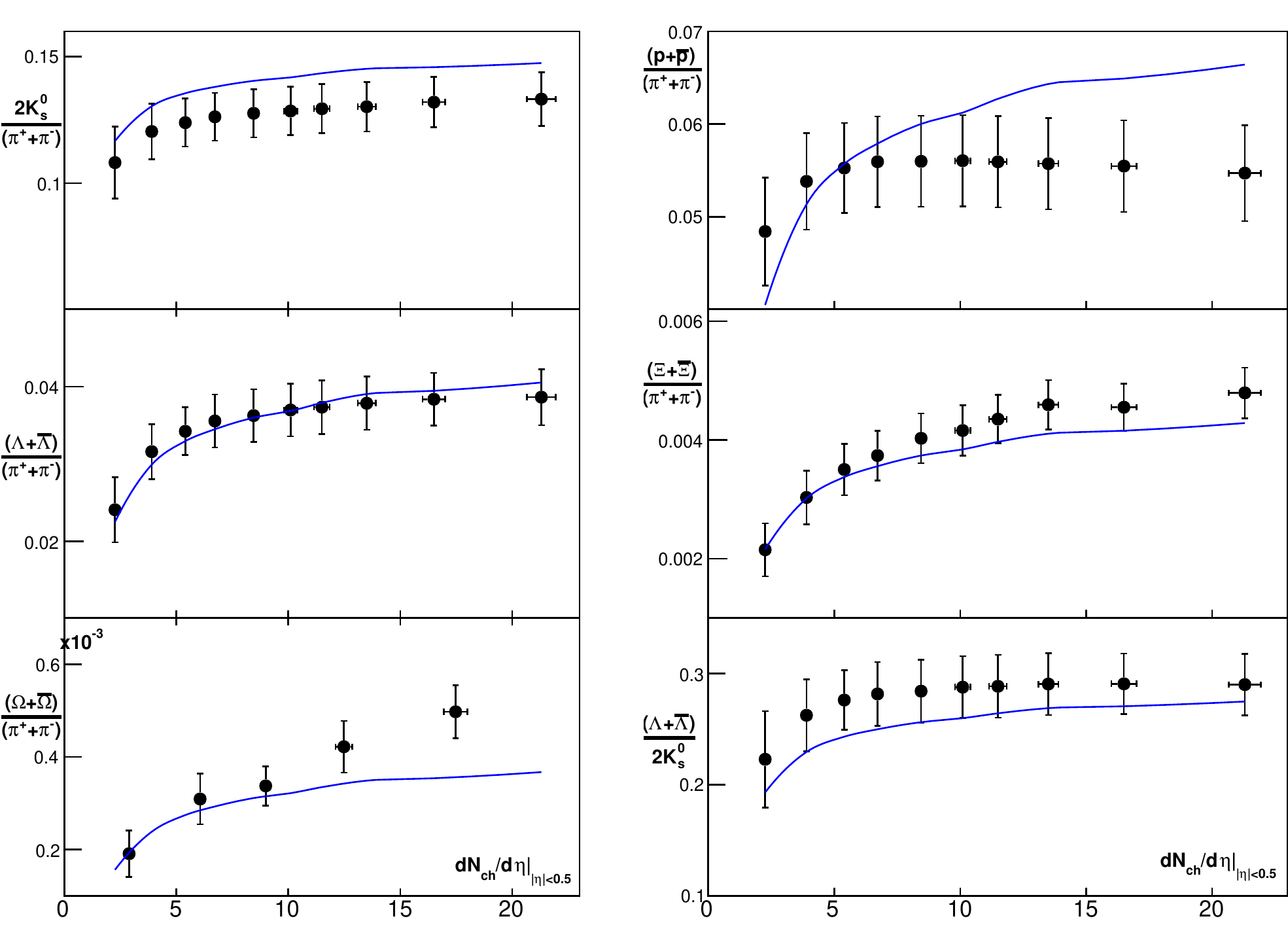}
\caption{ Different particle ratios as a function of $\dndetamid$ (symbols) \cite{ALICE:2017jyt} are compared to model calculations (line), Eq.(\ref{eq:ratio2}), performed at $Y_{FB1}$ values which assure minimum $\chi^2$ without including $\ppi$} \label{fig:fig4} } 
\end{figure}

To ensure the above result, a simplified picture was used to check whether the adjusted model can reproduce the measurements or not. Calculations at mid-rapidity in the two-fireball picture has an important feature: in every multiplicity class, there is a configuration - pair of fireballs rapidities- which has on average 75\% probability. 
This most probable configuration takes place when one fireball has negligible-contribution at $|y| \leq 0.5$. 
This feature will prominently hold for any reasonable $\psi(Y_{FB})$ \footnote{Reasonable $\psi(Y_{FB})$ is the one gives high probability for fireballs moving with high rapidities and producing a few number of particles.}. 
Based on that the second term between brackets in Eq.(\ref{eq:ratio1}) will be neglected and the equation becomes,
 \begin{equation} 
\label{eq:ratio2}
 (\frac{j}{i})_{X} = (\frac{j}{i})_{(fireball)} \frac{
  \int_{-0.5}^{0.5} \frac{dn_{j}}{dy} (Y_{FB1})~~dy}
 {\int_{-0.5}^{0.5} \frac{dn_{i}}{dy} (Y_{FB1})~~dy}
\end{equation}
In Eq.(\ref{eq:ratio2}), $Y_{FB1}$ is the only unknown parameter which can be extracted by fitting the  experimental particle ratios with the corresponding theoretical ones under the condition of minimizing  $\chi^2$ at each multiplicity class. 
The experimental ratios \cite{ALICE:2017jyt} are implemented as inputs, unlike Eq.(\ref{eq:ratio1}), to infer information about the system. 
Then, the validity of the model can be checked through the fit quality, $\chi^2/Ndf$, and the behavior of extracted parameter comparing with the expected one.
The statistically independent ratios, used in the fitting, are $\kspi$, $\ppi$, $\llpi$ and $\xipi$.
The fit has been performed in two ways with/without including $\ppi$.
The deduced $Y_{FB1}$ from both ways are listed in Tab.(\ref{tab:1}) while the experimental and corresponding theoretical results are illustrated in Fig.(\ref{fig:fig4}). In this figure, ratios of particle yields relative to pions are compared with model calculations in the case of excluding $\ppi$.
Few remarks about the fit and Fig.(\ref{fig:fig4}) are now in order,
\begin{itemize}
\item A remarkable agreement with particle ratios is seen except for $\ppi$. Moreover, the values $\chi^2/Ndf$ are less than the corresponding ones where SHM applied without any cuts effect especially the case in which $\ppi$ was excluded, as shown in Tab.(\ref{tab:1}).
\item The fit quality decreases in both cases as $\dndetamid$ increases. 
\item The extracted parameter, $Y_{FB1}$, decreases as $\dndetamid$ increases, in both cases, which is consistent with the main idea, Eq.(\ref{eq:main}). 
\item The model calculations agree clearly with the experimental ratios $\llks$ and $\omegapi$ which were excluded from the fit.     
\end{itemize}
  \begin{table}[htb]
\begin{tabular}{|c||   c    | c     || c |    c ||}     
\hline
 & \multicolumn{2}{|c||}{with $\ppi$} &\multicolumn{2}{|c||}{without $\ppi$}\\
\hline     
 $\dndetamid$ & $Y_{FB1}$  &  $\chi^2/Ndf$ &  $Y_{FB1}$  &  $\chi^2/Ndf$ \\ [0.5ex] 
\hline       
 21.29   &  0.5437       &       7.12/3          & 0.4973        &       3.52/2 \\
 16.51  &   0.5536      &       5.62/3          & 0.5177        &       3.03/2 \\
 13.46  &  0.5592       &       5.97/3          & 0.5255        &       3.85/2 \\
 11.51  &  0.5725       &       4.35/3          & 0.5464        &       3.01/2 \\
 10.08  &  0.5852       &       3.26/3          & 0.5656        &       2.49/2 \\
 8.45   &  0.596        &       2.77/3          & 0.5809        &       2.30/2 \\
 6.72   &  0.6144       &       1.95/3          & 0.6072        &       1.84/2 \\
 5.40   & 0.6353        &       1.76/3          & 0.6337        &       1.75/2 \\
 3.9    &  0.6766       &       1.21/3          & 0.6857        &       1.06/2 \\
 2.26   &  0.7829       &       1.75/3          & 0.8127        &       0.50/2 \\ [1ex]
\hline
\end{tabular}
\caption{The rapidities of the only fireball which emits in the measured region (see text), $Y_{FB1}$, are estimated from $\chi^2$ fitting with/without including $\ppi$ along with the minima $\chi^2/Ndf$ values.\label{tab:1}}
\end{table}
   It is remarkable that after the model has been reformed to match the experimental multiplicity cut,  it succeeded in reproducing the particle ratios mentioned in Fig.(\ref{fig:fig4}) except for $\ppi$ which has the largest deviation from SHM prediction at 2.76 TeV $\PbPb$ \cite{Andronic:2017pug}. Therefore,  if we assume that the puzzling behavior of $\ppi$ takes place in both Pb-Pb and high multiplicity pp events, the observed strangeness enhancement as a function of $\dndetamid$  would be a pure effect of the experimental cuts. Dealing with such kind of measurements, if necessary, required additional dynamical inputs, as shown in Eq.(\ref{eq:ratio1}) which was simplified and turned out to be Eq.(\ref{eq:ratio2}). This view provides a generalization of experimental cuts effect, specific form is suggested in Ref.\cite{Becattini:2003wp}. In that form, the rapidity cut could lead to artificial enhancement of strange particle assuming single static fireball. However, at LHC energies the cut at rapidity has no effect unless it is associated with another cut on multiplicity.
   
\section{conclusion}
\label{sec:con}
This work sheds a light on the consequence of the experimental cuts on rapidity and multiplicity. 
An approximation was used to establish a reasonable comparison with the experimental data.
The conclusion of such comparison is that if $\ppi$ is excluded and considered as an extension of abnormal behavior previously seen in Pb-Pb, the multiplicity dependence of particle ratios is nothing but an effect of a cut on charged particle multiplicity density associated with another cut on rapidity. 
In other words, the measured particle ratios under two simultaneous cuts on rapidity and multiplicity do not reflect the fireball hadronic composition. 
 It is possible to link results from different colliding systems using final state multiplicity size, if multiplicity cut will be the only  applied cut.
 Since, $\dndetamid$ becomes a widely used scaling, it is worthy to be deeply investigated. Further improvements are needed through adding more restrictions as well as extending to higher energies and $\pPb$.

\end{document}